\documentstyle[12pt]{article}
\title{ {\normalsize{{\hskip 9cm} BIHEP-TH-95-16, April 1995}}\\
        Space-like Penguin Effects in $B \rightarrow PP$ Decays }
\author{ Dong-Sheng Du$\rm ^{a,b}$, Mao-Zhi Yang$\rm ^b$,
         Deng-Zhi Zhang$ \rm ^{a,b}$
         \thanks{E-mail address: Duds@bepc3.ihep.ac.cn;$~~$
         Yangmz@bepc3.ihep.ac.cn; $~~$Zhangdz@bepc3.ihep.ac.cn} \\
         $\rm ^a$ \it{CCAST(World Laboratory),
         ~~P.O.Box 8730, Beijing 100080, China}\\
         $\rm ^b$ \it{Institute of High Energy Physics, Chinese
         Academy of Sciences,}\\
         \it{P.O.Box 918(4), Beijing, 100039, P. R. China}
         \thanks{mailing address.}}

\date{}
\begin{document}
\maketitle
\vspace*{0.3cm}
\begin{flushleft}
13.25.Hw, 13.40Hq
\end{flushleft}

\unitlength=1cm
\begin{flushleft}
\begin{picture}(16, 0.2)(0, 0)
\put(0, 0){\line(1, 0){16}}
\end{picture}
\section*{Abstract}
\end{flushleft}

The space-like penguin contributions to branching ratios and CP asymmetries
in charmless decays of $B$ to two pseudoscalar mesons are studied using the
next-to-leading order low energy effective Hamiltonian. Both the gluonic
penguin and the electroweak penguin diagrams are considered. We find that
the effects are significant.

\begin{flushleft}
\begin{picture}(16, 0.2)(0, 0)
\put(0, 0){\line(1, 0){16}}
\end{picture}
\end{flushleft}

\newpage
\begin{flushleft}
\section*{1. Introduction}
\end{flushleft}

Penguin diagrams play an important role in charmless B-decays and direct
CP violation [1,2]. But only time-like penguin diagrams were considered in
the literature because they can provide the necessary different strong
phases for CP violation by different loop effects of the internal $u$ and $c$
quarks [1]. The contribution of space-like penguin diagrams is usually
neglected assuming form factor suppression. This assumption for neglecting
space-like penguin effects is used not only for gluonic penguins but also for
electro-weak penguins [3]. But it does not lie on a solid ground because
the space-like penguin amplitudes can be remarkably enhanced by the hadronic
matrix elements involving (V-A)(V+A) or (S+P)(S-P) currents [4]. Although
space-like penguin diagrams can only provide an overall CP conserving
phase due to final state interaction, it affects CP asymmetry by modifying
the dispersive or absorptive parts of time-like penguin amplitudes. Obviously,
it affects branching fractions too. In our recent paper [5], we illustrated
the space-like penguin effects in CP asymmetries for the exclusive B-decays
$B_u^-\to \bar{K^0}\pi^-$ and $K^0K^-$ using leading order Hamiltonian.
In contrast to the naive expectation, the space-like penguin effects on CP
asymmetries are found to be significant. In this letter we study space-like
penguin effects in B to two pseudoscalar decays systematically.
We concentrate on
the charmless B decays because penguins play an important role in these
decays. We use the next to leading order low energy effective Hamiltonian
in order to consider both gluonic and electro-weak penguins. We can see
later that the contribution of the electro-weak penguin is not negligible.
The article is organized as following: In section II, We present the next
to leading order effective Hamiltonian and the computation method. Section
III devotes to the numerical results and corresponding discussions.

\begin{flushleft}
\section*{2. Effective hamiltonian and factorization approximation}
\end{flushleft}

Following ref. [6], the next-to-leading order low energy effective
Hamiltonian describing $\Delta B=-1$, $\Delta C=\Delta U=0$ transitions
is given at the renormalization scale $\mu=O(m_b)$ as
$$
{\cal H}_{eff}(\Delta B=-1) = \frac{G_F}{\sqrt{2}}
\left[\sum_{q=u,c}v_q \left\{ Q_1^qC_1(\mu)+Q_2^qC_2(\mu)
                       +\sum_{k=3}^{10} Q_k C_k(\mu)\right\}\right],
\eqno(1)
$$
where the Wilson coefficient functions $C_i(\mu)$ (i=1,$\cdots$,10)
are calculated in renormalization group improved perturbation theory and
include leading and next-to-leading order QCD corrections and leading order
corrections in $\alpha$. The CKM factors $v_q$ are defined as
$$
v_q=\left\{
          \begin{array}{ll}
          V_{qd}^*V_{qb} & \mbox{for $b \rightarrow d$ transitions}\\
          V_{qs}^*V_{qb} & \mbox{for $b \rightarrow s$ transitions.}
          \end{array}
     \right.
\eqno(2)
$$
Here, we make use of the Wolfenstein parametrization[7] in which the CKM
matrix can be written in terms of four parameters $\lambda$, $A$,
$\rho$ and $\eta$ in the following form:
$$
V=\left|
\begin{array}{ccc}
1-\frac{1}{2}\lambda^2 & \lambda & \lambda^3 A(\rho-i\eta)\\
  -\lambda   &         1-\frac{1}{2}\lambda^2  &  \lambda^2 A\\
\lambda^3 A(1-\rho-i\eta)  &     -\lambda^2 A    &             1
\end{array}
\right|.
\eqno(3)
$$
The preferred values of the CKM Parameters are $\lambda=0.22$,
$A=0.8$, $\eta=0.34$, $\rho=-0.12$, which are obtained from the fit to
experimental data[8].
The operators $Q_1^u$, $Q_2^u$, $Q_3,\ldots,Q_{10}$ are given as the
following forms:
$$
\begin{array}{rlrl}
Q_1^u= & (\bar{q}_{\alpha}u_{\beta})_{V-A}(\bar{u}_{\beta}b_{\alpha})_{V-A} &
Q_2^u= & (\bar{q}u)_{V-A}(\bar{u}b)_{V-A}\\
Q_3= & (\bar{q}b)_{V-A}\displaystyle\sum_{q'}(\bar{q}'q')_{V-A} &
Q_4= & (\bar{q}_{\alpha}b_{\beta})_{V-A}
      \displaystyle\sum_{q'}(\bar{q}'_{\beta}q'_{\alpha})_{V-A} \\
Q_5= & (\bar{q}b)_{V-A}
      \displaystyle\sum_{q'}(\bar{q}'q')_{V+A} &
Q_6= & (\bar{q}_{\alpha}b_{\beta})_{V-A}
      \displaystyle\sum_{q'}(\bar{q}'_{\beta}q'_{\alpha})_{V+A} \\
Q_7= & \frac{3}{2}(\bar{q}b)_{V-A}
      \displaystyle\sum_{q'}e_{q'}(\bar{q}'q')_{V+A} &
Q_8= & \frac{3}{2}(\bar{q}_{\alpha}b_{\beta})_{V-A}
      \displaystyle\sum_{q'}e_{q'}(\bar{q}'_{\beta}q'_{\alpha})_{V+A} \\
Q_9= & \frac{3}{2}(\bar{q}b)_{V-A}
      \displaystyle\sum_{q'}e_{q'}(\bar{q}'q')_{V-A} &
Q_{10}= & \frac{3}{2}(\bar{q}_{\alpha}q_{\beta})_{V-A}
      \displaystyle\sum_{q'}e_{q'}(\bar{q}'_{\beta}q'_{\alpha})_{V-A}.
\end{array}
\eqno(4)
$$
where $Q_1^u$ and $Q_2^u$ are the current-current operators, and
the current-current operators $Q_1^c$ and $Q_2^c$ are obtained from
$Q_1^u$ and $Q_2^u$ through the substitution of $u\rightarrow c$.
$Q_3,\ldots,Q_6$ are the QCD penguin operators, whereas $Q_7,\ldots,Q_{10}$
are the electroweak penguin operators. The quark $q=d~or~s$ is for
$b \rightarrow
d~or~s$ transitions, respectively; $q'$ is running over the quark
flavours being active at the scale $\mu=O(m_b)~(q'\in \{u, d, c, s, b\})$;
$e_{q'}$ are the corresponding quark charges; the indices $\alpha,~\beta$
are $SU(3)_c$ color indices; $(V \pm A)$ refer to $\gamma_{\mu}(1\pm
\gamma_5)$. It should be noted that the Hamiltonian (1) can be viewed as
the generalization of the leading logarithmic Hamiltonians presented in
[9,10].

Beyond the leading logarithmic approximation, the Wilson coefficient
functions $C_i(\mu)$ depend both on the form
of the operator basis (4) and on the renormalization scheme. Here,
we use the renormalization scheme independent Wilson coefficient functions
[11]:
$$
{\bf \bar{C}}(\mu)=\left[\hat{1}+\frac{\alpha_s(\mu)}{4\pi}\hat{r}_s^T
                    +\frac{\alpha(\mu)}{4\pi}\hat{r}_e^T\right]
                    \cdot {\bf C}(\mu),
\eqno(5)
$$
where $\hat{r}_s$ and $\hat{r}_e$ are obtained from
one-loop matching conditions. Now, taking the QCD and
electroweak one-loop level matrix elements of the operators $Q_i$
($Q_i$=$Q_1^u$, $Q_2^u$, $Q_3,\ldots,Q_{10}$) into account through
$$
<{\bf Q}^T(\mu)>=<{\bf Q}^T>_0\cdot \left[\hat{1}+\frac{\alpha_s(\mu)}{4\pi}
 \hat{m}^T_s(\mu)+\frac{\alpha_{em}}{4\pi}\hat{m}^T_e(\mu)\right],
\eqno(6)
$$
which define matrices $\hat{m}_s(\mu)$ and $\hat{m}_e(\mu)$.
In Eq. (5) and (6), ${\bf C}(\mu)$, ${\bf \bar{C}}(\mu)$ and ${\bf Q}$
are all column vectors, where the vector ${\bf Q}$ are given by the operator
basis $Q_i$, and $<{\bf Q}>_0$ denote
the tree level matrix elements of these operators. Combine Eq.(5) and (6),
we obtain
$$
\begin{array}{rl}
<{\bf Q}^T(\mu)\cdot {\bf C}(\mu)>  \\
&=~<{\bf Q}^T>_0\cdot \left[\hat{1}+
\displaystyle\frac{\alpha_s(\mu)}{4\pi}\left(\hat{m}_s(\mu)-\hat{r}_s\right)^T+
\displaystyle\frac{\alpha(\mu)}{4\pi}\left(\hat{m}_e(\mu)-\hat{r}_e\right)^T
\right]\cdot {\bf \bar{C}}(\mu)  \\
&\equiv~<{\bf Q}^T>_0\cdot{\bf C'}(\mu)
\end{array} \eqno(7)
$$
where ${\bf C'}(\mu)$ are defined as
$$
\begin{array}{rl}
C'_1&=~\overline{C}_1,~~~~~~~~~C'_2~=~\overline{C}_2,\\
C'_3&=~\overline{C}_3-P_s/3,~~~~C'_4~=~\overline{C}_4+P_s,\\
C'_5&=~\overline{C}_5-P_s/3,~~~~C'_6~=~\overline{C}_6+P_s,\\
C'_7&=~\overline{C}_7+P_e,~~~~~~C'_8~=~\overline{C}_8,\\
C'_9&=~\overline{C}_9+P_e,~~~~~~C'_{10}~=~\overline{C}_{10},
\end{array} \eqno(8)
$$
where $P_{s,e}$ are given by
$$
\begin{array}{rl}
P_s&=~\frac{\alpha_s}{8\pi}\overline{C}_2(\mu)\left[\frac{10}{9}-
       G(m_q,q,\mu)\right],\\
P_e&=~\frac{\alpha_{em}}{9\pi}\left(3\overline{C}_1+\overline{C}_2(\mu)
      \right)\left[\frac{10}{9}-G(m_q,q,\mu)\right],\\
G(m,q,\mu)&=~-4\int_0^1 dx~x(1-x)ln\displaystyle\left[\frac{m^2-x(1-x)q^2}
{\mu^2}\right],
\end{array} \eqno(9)
$$
here $q=u,~c$, for numerical calculation, we take $m_u=0.005GeV$,
$m_c=1.35GeV$, and ${\sl q^2}$ denotes the momentum transfer squared
of the virtual gluons,
photons and $Z^0$ appearing in the QCD and electroweak penguin matrix elements.
For the details of this calculation, the reader is referred to ref. [12,13].

The renormalization scheme independent Wilson coefficient functions
${\bar C}_i(\mu)$ at the scale $\mu=O(m_b)$ are obtained by first calculating
the Wilson coefficients at $\mu=O(m_W)$ and then using the renormalization
group equation to evolve them down to $O(m_b)$. We use in our analysis,
$\alpha_s(m_Z)=0.118$, $\alpha(m_Z)=1/128$[14] and $m_t=174$GeV[15]
and the numerical values of the renormalization scheme
independent Wilson Coefficients $\overline{C}_i(\mu)$ at $\mu=O(m_b)$
are[13]
$$
\begin{array}{rl}
\bar{ c}_1&=~-0.313, ~~ \bar {c}_2=1.150,~~\bar{ c}_3=0.017,\\
\bar{ c}_4&=~-0.037,~~ \bar {c}_5=0.010, ~~ \bar{ c}_6=-0.046,\\
\bar{ c}_7&=~-0.001\cdot \alpha_{em}, ~~ \bar {c}_8=0.049\cdot\alpha_{em},
{}~~ \bar{ c}_9=-1.321\cdot\alpha_{em}, \\
\bar{ c}_{10}&=~0.267\cdot\alpha_{em}.
\end{array}\eqno(10)
$$

With the help of Eq. (7), the two-body decay amplitude $<PP^{'}|H_{eff}(\Delta
B=-1)|B>$ can be expressed as linear combinations of $<PP^{'}|Q_i|B>_0$.
The hadronic matrix elements $<PP^{'}|Q_i|B>_0$ are
evaluated using the factorization approximation [16].
It should be noted that this approach has already been
used in the literature to analyze the QCD or electroweak time-like penguin
contributions[12].
However, we go further in this letter by including
the space-like penguin diagrams.  As in [5,12], we also
neglect W-annihilation or W-exchange diagram contributions in our present
analysis which are commonly assumed to be form factor suppressed.

Using the vacuum-saturation approximation, the decay amplitude
$<PP'|H_{eff}|B>$ can be factorized into a product of two current matrix
elements $<P|J^{\mu}|0>$ and $<P'|J'_{\mu}|B>$ for the tree and time-like
penguin diagrams (Fig.1), or the product of $<pp'|J^{\mu}|0>$ and
$<0|J'_{\mu}|B>$ for the space-like penguin diagrams (Fig.2). In this work,
the hadronic matrix elements are calculated in BSW method[16]. We define
$$
\begin{array}{rl}
M^{pp'}_{q_1q_2q_3}&=~<P|(\bar {q_1}q_2)_{V-A}|0><P'|(\bar {b}q_3)_{V-A}|B>,\\
              &\stackrel{or}{=}~<P|(\bar {q_1}q_2)_{V-A}|0><P'|(\bar{
q_3}b)_{V-A}|B>,
\end{array} \eqno(11)
$$
and
$$
\begin{array}{rl}
S^{pp'}_{q_1q_2q_3}&=~<PP'|(\bar{ q_1}q_2)_{V-A}|0><0|(\bar{ b}q_3)_{V-A}|B>,\\
              &\stackrel{or}{=}~<PP'|(\bar{ q_1}q_2)_{V-A}|0><0|(\bar
{q_3}b)_{V-A}|B>,
\end{array} \eqno(12)
$$
where $M^{pp'}_{q_1q_2q_3}$ denotes the hadronic matrix element in tree
and time-like penguin diagram case, while $S^{pp'}_{q_1q_2q_3}$ denotes
space-like penguin case. When the (V-A)(V+A) current are transformed
into (S+P)(S-P) and further into (V-A)(V-A) ones using equation of motion
for the time-like and space-like penguin amplitudes,
there appear the terms which are proportional to
$\displaystyle\frac{2m^2_X}{(m_q+m_{q'})(m_b-m_{q'})}$ and
$\displaystyle\frac{2m^2_B}{(m_q-m_{q'})(m_b+m_{q'})}$, respectively.
If $q=q'$ as in the decay modes:
$$
\begin{array}{rll}
\mbox{$\bar{B_d^0} \rightarrow$} &
\mbox{$\pi^- \pi^+$, $\pi^0 \pi^0$, $\pi^0 \eta$, $\pi^0 \eta^{'}$,
  $\eta \eta$, $\eta \eta^{'}$, $\eta^{'} \eta^{'}$, $K^0 \bar{K^0}$,
  ;} &
\mbox{(for $b \rightarrow d$ transitions)}\\
\mbox{$\bar{B_s^0} \rightarrow$} &
\mbox{$K^- K^+$, $\bar{K^0} K^0$, $\pi^0 \eta$, $\pi^0 \eta^{'}$,
  $\eta \eta$, $\eta \eta^{'}$, $\eta^{'} \eta^{'}$.} &
\mbox{(for $b \rightarrow s$ transitions),}
\end{array}
$$
the denominator of the factor
$\displaystyle\frac{2m^2_B}{(m_q-m_{q'})(m_b+m_{q'})}$
is zero. So, we can not use equation of motion
to compute the amplitudes of these decays. We have to compute the matrix
elements of (S+P)(S-P) operators directly. We shall discuss it elsewhere.

As an example of how to factorize the decay amplitudes into the product
of hadronic matrix elements, we give the result of
$<\pi^-\pi^0|H_{eff}|B_u^->$ in the following,
$$\begin{array}{rl}
&<\pi^-\pi^0|H_{eff}|B_u^->~~~~~~~~~~~~~~~~~~~~~~~~~~~~~~~~~~~~~~~~~~~~~~~~\\
=&\frac{G_F}{\sqrt{2}}\displaystyle\sum_{q=u,c} v_q\left[\{a_1\delta_{uq}
+a_3-\displaystyle\frac{2M_{\pi^-}^2}{(m_d+m_u)(m_u-m_b)}(a_5+a_7)+a_9\}
M^{\pi^-\pi^0}_{duu}+\right.\\[4mm]
&+\{a_2\delta_{uq}-a_3+\displaystyle\frac{M_{\pi^0}^2}{m_d(m_d-m_b)}
 (a_5-a_7/2)
-\frac{3}{2}a_8+\frac{3}{2}a_{10}+\frac{1}{2}a_9\}
M^{\pi^0\pi^-}_{uud}+\\[4mm]
&+\left.\{a_3+\displaystyle\frac{2M_{B}^2}{(m_u+m_b)(m_d-m_u)}(a_5+a_7)+a_9\}
(S^{\pi^-\pi^0}_{duu}+S^{\pi^0\pi^-}_{duu})\right],
\end{array}\eqno(13)
$$
where the term $(S^{\pi^-\pi^0}_{duu}+S^{\pi^0\pi^-}_{duu})$ is the
contribution obtained from two space-like penguin diagrams, and the quark
masses are taken as $m_d=0.01GeV$, $m_u=0.005GeV$, $m_s=0.175GeV$,
$m_b=5.0GeV$. $a_k$ is
defined as
$$\begin{array}{rl}
a_{2i-1}&\equiv~\displaystyle\frac{C'_{2i-1}}{3}+C'_{2i},\\[4mm]
a_{2i}&\equiv~ C'_{2i-1}+\displaystyle\frac{C'_{2i}}{3},~(i=1,2,3,4,5)
\end{array}
$$
The general expression for the one-body pseudoscalar matrix element
of the axial-vector is
$$
<0|V_{\mu}-A_{\mu}|P(q)>=if_Pq_\mu,
\eqno(14)
$$
where q represents the momentum of the pseudoscalar meson, and
$f_P$ is the decay constant. The two-body pseudoscalar-pseudoscalar
matrix element of the vector current is[4, 17]
$$
<P_2(q_2)|V_{\mu}-A_{\mu}|P_1(q_1)>=f_+(q_-^2)q_{+\mu}+f_-(q_-^2)q_{-\mu},
\eqno(15)
$$
where $q_\pm=q_1 \pm q_2$, and the form factor $f_{\pm}$ is given by
the monopole parametrization
$$
f_+(q_-^2)\simeq \displaystyle\frac{f_+(0)}{1-q^2/m^2_{pole}}
\eqno(16\rm a)
$$
$$
f_-(q_-^2)=-\displaystyle\frac{m_1-m_2}{m_1+m_2}f_+(q_-^2).
\eqno(16\rm b)
$$

With equation (14$\sim$16), we obtain
$$\begin{array}{rl}
M_{duu}^{\pi^-\pi^0}&=~-\frac{i}{\sqrt{2}}f_{\pi^-}f_+^{B_u^-\pi^0}(M_{\pi^-}^2)
    \displaystyle\frac{M_{B_u^-}-M_{\pi^0}}{M_{B_u^-}+M_{\pi^0}}\left[
   (M_{B_u^-}+M_{\pi^0})^2-M_{\pi^-}^2\right],\\[4mm]
M_{uud}^{\pi^0\pi^-}&=~-if_{\pi^0}^{\bar {u}u}f_+^{B_u^-\pi^-}(M_{\pi^-}^2)
    \displaystyle\frac{M_{B_u^-}-M_{\pi^-}}{M_{B_u^-}+M_{\pi^-}}\left[
   (M_{B_u^-}+M_{\pi^-})^2-M_{\pi^0}^2\right],
\end{array}\eqno(17)
$$
and
$$\begin{array}{rl}
S_{duu}^{\pi^-\pi^0}&=~\frac{i}{\sqrt{2}}f_{B_u^-}f_+^a(M_{B_u^-}^2)
    \displaystyle\frac{M_{\pi^-}-M_{\pi^0}}{M_{\pi^-}+M_{\pi^0}}\left[
   (M_{\pi^-}+M_{\pi^0})^2-M_{B_u^-}^2\right],\\[4mm]
S_{duu}^{\pi^0\pi^-}&=~-\frac{i}{\sqrt{2}}f_{B_u^-}f_+^a(M_{B_u^-}^2)
    \displaystyle\frac{M_{\pi^0}-M_{\pi^-}}{M_{\pi^-}+M_{\pi^0}}\left[
   (M_{\pi^-}+M_{\pi^0})^2-M_{B_u^-}^2\right],
\end{array} \eqno(18)
$$
where the factors $\frac{1}{\sqrt{2}}$, $-\frac{1}{\sqrt{2}}$ come
from the constituent
of $\pi^0=\frac{1}{\sqrt{2}}(\bar{ u}u-\bar{ d}d)$.

In order to give numerical results, we need to know the form factors. For the
decay form factors like $f_+^{B\pi}(M^2)$, etc., we can use BSW [16] method
to calculate them. For the annihilation form factor $f_+^a(Q^2)$, we do not
have reliable method to compute it. But at $Q^2=M_B^2$ one is far from the
$K\pi$, $\pi\pi$, $\eta K$ resonance region. So, for the charmless B decays,
because the large energy release, we can use the form factor in its asymptotic
form. For charmless B to two pseudoscalars decays, the asymptotic form factor
predicted by QCD [18] should be a resonable approximation. So we take
$f_+^a(Q^2)=i 16\pi\alpha_s f_B^2/Q^2$. Now we are in a position to give the
numerical results.
\begin{flushleft}
\section*{3. Results and discussions}
\end{flushleft}

The decay width of a B-meson at rest decaying into two pseudoscalars is
$$
\Gamma(B \to PP^{'})=\frac{1}{8\pi}|<PP^{'}|H_{eff}|B>|^2\frac{|p|}{M^2_B},
\eqno(19)
$$
where
$$
|p|=\frac{[(M_B^2-(M_P+M_P{'})^2)(M_B^2-(M_P-M_P{'})^2)]^{\frac{1}{2}}}{2M_B}
\eqno(20)
$$
is the momentum of the pseudoscalar meson $P$ or $P{'}$.
The corresponding branching ratios are given by
$$
Br(B\to PP^{'})=\frac{\Gamma(B \rightarrow PP^{'})}{\Gamma_{tot}}.
\eqno(21)
$$
In our numerical calculation, we take[14]
$\Gamma_{tot}^{B_u^-}=4.27\times10^{-13}$GeV,
$\Gamma_{tot}^{B_d^0}=4.39\times10^{-13}$GeV, and
$\Gamma_{tot}^{B_s^0}=4.91\times10^{-13}$GeV.

In order to obtain the CP-violating parameter, the B-meson decay amplitude
can be generally expressed as
$$
\begin{array}{rl}
<PP'|H_{eff}|B>&=~\displaystyle\frac{G_F}{\sqrt{2}}
               \displaystyle\sum_{q=u,c}v_q(C'_1<Q_1^q>+C'_2<Q_2^q>\\
               &~~+\displaystyle\sum_{k=3}^{10}C'_k<Q_k>)\\
               &\equiv\displaystyle\frac{G_F}{\sqrt{2}}
                \displaystyle\sum_{q=u,c}v_q F_q.
\end{array} \eqno(22)
$$
 With the help of eq.(22), one can get the CP-violating asymmetry parameter
$$\begin{array}{rl}
\displaystyle{\cal A}_{cp}
&=~\displaystyle\frac{\Gamma(B \rightarrow PP^{'})-
       \Gamma({\bar B} \rightarrow \bar{P}\bar{P{'}})}
       {\Gamma(B \rightarrow PP^{'})+
       \Gamma({\bar B} \rightarrow \bar{P}\bar{P{'}})}\\[4mm]
&=~\displaystyle\frac{2Im(v_uv_c^*)Im(\frac{F_c}{F_u})}{v_u^2+
    v_c^2(\frac{F_c}{F_u})^2+2Re(\frac{F_c}{F_u})}.
\end{array} \eqno(23)
$$

Since the branching ratios and CP asymmetries depend crucially on the
parameter $q^2$ describing the momentum squared of
the exchanged virtual particles
appearing in the penguin matrix elements of Fig. 1 and 2, we should
consider it in detail. Here, we use the same simple picture for two-body
decays illustrated in Fig. 1c and 2 as the one in ref [5].
With the simple physical picture presented in Ref.[5], the average
value of the momentum squared $<q^2>$ of the exchanged virtual particles can
be given by
$$
<q^{2}>=m^{2}_{b}+m^{2}_{q}-2m_{b}E_{q}~,
\eqno(24)
$$
where $E_{q}$ is determined from
$$
E_{q}+\sqrt{E^{2}_{q}-m^{2}_{q}+m^{2}_{q^{'}}}
+\sqrt{4E^{2}_{q}-4m^{2}_{q}+m^{2}_{q^{'}}}=m_{b};
\eqno(25{\rm a})
$$
for the time-like penguin channels; or from
$$
E_{q}+\sqrt{E^{2}_{q}-m^{2}_{q}+m^{2}_{q^{'}}}
=m_{b}+m_{q^{'}};
\eqno(25{\rm b})
$$
for the space-like penguin channels. When we factorize $<Q_k>_0$ of hairpin
diagrams illustrated in Fig. 1d), we find $<Q_3>_0=-<Q_5>_0$,
$<Q_4>_0=-<Q_6>_0$, and $<Q_7>_0=-<Q_9>_0$. and hence the factor in Eq. 6:
$$
\begin{array}{l}
\left\{
  \displaystyle\frac{\alpha_s}{8\pi}
  \left[-\displaystyle\frac{1}{N_c}<Q_3>_0+<Q_4>_0-
  \displaystyle\frac{1}{N_c}<Q_5>_0+<Q_6>_0
  \right]\bar{C}_2(\mu)
\right.\\
\left.
+\displaystyle\frac{\alpha}{3\pi}
  \left[<Q_7>_0+<Q_9>_0\right]\left[\bar{C}_1(\mu)
    +\displaystyle\frac{1}{N_c}\bar{C}_2(\mu)\right]
\right\}
\end{array}
\eqno(26)
$$
vanishes because of the cancelation.
So, we do not need to consider the hairpin diagrams.

The numerical results of the space-like penguin contributions to the
branching ratios and CP-violating asymmetries are given in table 1.
In the meantime, we also calculate the branching ratios and
CP-violating asymmetries with only the tree and time-like penguin
contributions for comparison. We also present the results with only tree and
gluonic penguin contributions. All the parameters such as meson decay constants
and form factors needed in our calculation are taken as
$f_{\pi^{\pm}}=0.13GeV$,
$f_K=0.160GeV$[14], $f_{\pi^0}^{{\bar u}u}=-f_{\pi^0}^{\bar {d}d}=f_{\pi^\pm}
/\sqrt{2}$. $f_{\eta}^{\bar {u}u}=f_{\eta}^{\bar {d}d}=0.092$,
$f_{\eta}^{\bar {s}s}=-0.105$,
$f_{\eta^{'}}^{\bar {u}u}=f_{\eta^{'}}^{\bar {d}d}=0.049$,
$f_{\eta{'}}^{\bar {s}s}=0.12$[4],
$f_D=0.23$, $f_{D_s}=0.281$[19],
$f_B=1.5\times f_{\pi^{\pm}}$[20], $f_{B_s}=0.206$[21],and
$f_+^{B_u^-\pi^-}(0)=0.29$, $f_+^{B_u^-K^-}(0)=0.32$[22],
$f_+^{B_u^-\eta_{\bar {u}u}}(0)=0.307$,
$f_+^{B_u^-\eta^{'}_{\bar {u}u}}(0)=0.254$,
$f_+^{BD}(0)=0.690$[16],
$f_+^{B_s^-\eta_{\bar {s}s}}(0)=0.335$,
$f_+^{B_s^-\eta^{'}_{\bar {s}s}}(0)=0.282$[23],
$f_+^{B_sD_s}(0)=0.648$[24].

{}From Table 1 we can see the following features:

(i) For most of the charmless decays, penguin contributions are important.

(ii) For $B_u^-\to\pi^-\eta$, $K^-\pi^0$, $K^-\eta$, $K^-\eta'$, $\bar{B_d^0}
\to\bar{K^0}\pi^0$, $\bar{K^0}\eta$, $\bar{K^0}\eta'$ and $B_s^0\to\pi^0 K^0$,
the contribution of the electro-weak penguins are not negligible.

(iii) The space-like penguin effects in $B_u^-\to \pi^-\pi^0$ are amazingly
large. The correction to the branching ratio is more than 100\%, while to
the CP asymmetry is about an order of magnitude, actually, $A_{cp}\sim 0.77\%$
with only time-like penguin, but $A_{cp}\sim-70.4\%$ when including the
space-like penguin.

For $B_u^-\to K^-\pi^0$, $K^-\eta$, $K^-\eta'$,
$\bar{K^0}\pi^-$, $\bar{B_d^0}\to K^-\pi^+$, $\bar{K^0}\eta$,$\bar{K^0}\eta'$,
and $\bar{B_s^0}\to \pi^- K^+$, $\pi^0 K^0$, the space-like penguin
contributions are also dominant.

In $B_u^-\to\pi^-\eta$, $\pi^-\eta'$, the space-like penguin contribution is
zero. The reason is that there are two space-like penguin diagrams in each
channel and the contributions of the two diagrams exactly cancel each other.

In general, we can conclude that the space-like penguin effects are not
negligible in most of the charmless two-pseudoscalar decays of the B mesons.
The space-like penguins can affect not only CP asymmetries, but also decay
branching ratios. Further investigations are definitely needed.

\begin{flushleft}
\section*{Acknoledgement}
\end{flushleft}

This work is supported in part by the China National Natural
Science Foundation and the Grant of State Commission of Science and
Technology of China.

\newpage

Table 1. The branching ratios and the CP asymmetries, where the ``Only Tree"
means the branching ratios with only  tree diagram contribution, ``T-like"
denotes the time-like penguin contributions, the ``S-like"
denotes the space-like penguin contributions, ``QCD" means QCD penguin and
tree diagrams contributions, and ``QCD+EW" denotes full tree, QCD and
EW(electro-weak) penguin contributions.
\newpage

\begin{center}
\begin{tabular}{|c|c|c|c|c|c|c|c|c|c|}
\hline
decay mode & \multicolumn{5}{|c|}{Br}
&\multicolumn{4}{|c|}{${\cal{A}}_{cp}$}\\
\cline{2-10} & Only Tree&\multicolumn{2}{|c|}{Tree+T-like} &
                   \multicolumn{2}{|c|}{Tree+T-like+S-like} &
                    \multicolumn{2}{|c|}{Tree+T-like} &
                   \multicolumn{2}{|c|}{Tree+T-like+S-like} \\
\cline{3-10} & &QCD& QCD+EW & QCD & QCD+EW & QCD & QCD+EW & QCD & QCD+EW \\
\hline
$B_u^- \rightarrow \pi^- \pi^0$     &$2.96\times 10^{-6}$&$2.90\times 10^{-6}$
     &$2.74\times 10^{-6}$ &$1.30\times 10^{-6}$&$1.12\times 10^{-6}$
        &$0.69\%$& $0.77\%$ &$-67.4\%$& $-70.4\%$ \\
\hline
$B_u^- \rightarrow \pi^- \eta$    &$2.19\times 10^{-6}$ &$3.68\times10^{-6}$
      & $2.81\times 10^{-6}$ &$3.68\times10^{-6}$ &$2.81\times 10^{-6}$
      &$35.1\%$ & $33.5\%$ &$35.1\%$&$33.5\%$ \\
\hline
$B_u^- \rightarrow \pi^- \eta{'}$  &$6.95\times 10^{-7}$ &$5.84\times10^{-6}$
       &$5.8\times 10^{-6}$ &$5.84\times10^{-6}$&$5.8\times 10^{-6}$
       &$16.0\%$& $16.1\%$ &$16.0\%$&$16.1\%$ \\
\hline
$B_u^- \rightarrow K^0 K^-$    &0 &$4.16\times10^{-7}$
      &$4.08\times 10^{-7}$ &$3.79\times10^{-7}$ &$3.72\times 10^{-7}$
      &$2.47\%$ & $2.49\%$ &$-2.21\%$& $-2.24\%$ \\
\hline
$B_u^- \rightarrow K^- \pi^0$  &$2.12\times 10^{-7}$   &$2.93\times10^{-6}$
       & $4.22\times 10^{-6}$ &$5.29\times10^{-6}$&$6.55\times 10^{-6}$
       &$-8.58\%$& $-6.15\%$ &$22.8\%$& $17.6\%$ \\
\hline
$B_u^- \rightarrow K^- \eta$   &$1.57\times 10^{-7}$   &$1.53\times10^{-7}$
      & $1.84\times 10^{-7}$ &$4.00\times10^{-8}$&$7.13\times 10^{-8}$
      &$6.17\%$ & $4.38\%$ &$-77.6\%$& $-66.1\%$ \\
\hline
$B_u^- \rightarrow K^- \eta^{'}$ &$4.98\times 10^{-8}$ &$8.19\times10^{-6}$
      &$7.74\times 10^{-6}$ &$8.03\times10^{-6}$&$7.59\times 10^{-6}$
      &$-3.35\%$ & $-3.53\%$ &$-1.09\%$& $-1.18\%$ \\
\hline
$B_u^- \rightarrow \bar{K^0} \pi^-$ &0 &$4.95\times10^{-6}$
      &$4.86\times 10^{-6}$ &$7.10\times10^{-6}$&$6.97\times 10^{-6}$
      &$-0.18\%$ & $-0.18\%$ &$0.87\%$&$ 0.88\%$ \\
\hline
$\bar{B_d^0} \rightarrow K^-\pi^+$ &$3.77\times 10^{-7}$ &$5.78\times10^{-6}$
      &$ 5.96\times 10^{-6}$ &$1.04\times10^{-5}$& $1.06\times 10^{-5}$
      &$-8.23\%$ & $-8.0\%$ &$21.8\%$ & $21.3\%$ \\
\hline
$\bar{B_d^0} \to\bar{K^0} \pi^0 $  &$6.76\times 10^{-10}$ &$2.46\times10^{-6}$
      &$1.56\times 10^{-6}$ &$3.64\times10^{-6}$&$2.74\times 10^{-6}$
      &$0.44\%$ & $0.73\%$  &$-0.74\%$ &$ -1.16\%$ \\
\hline
$\bar{B_d^0} \to \bar{K^0} \eta$ &$6.72\times 10^{-10}$ &$4.25\times10^{-8}$
      & $1.97\times 10^{-9}$ &$6.42\times10^{-8}$&$2.33\times 10^{-8}$
      &$1.40\%$  & $47.0\%$  &$-11.2\%$& $-26.9\%$ \\
\hline
$\bar{B_d^0} \rightarrow \bar{K^0} \eta^{'}$ &$1.87\times 10^{-10}$
&$7.97\times10^{-6}$
      & $7.42\times 10^{-6}$ &$7.64\times10^{-6}$&$7.09\times 10^{-6}$
      &$-0.42\%$  &$ -0.44\%$ &$-0.06\%$ &$ -0.068\%$ \\
\hline
$\bar{B_s^0} \rightarrow \pi^- K^+$ &$ 3.51\times 10^{-6}$ &$3.08\times10^{-6}$
        &$3.07\times 10^{-6}$  &$1.70\times10^{-6}$&$1.68\times 10^{-6}$
        &$8.34\%$& $8.41\%$ &$-44.1\%$ &$ -44.4\%$ \\
\hline
$\bar{B_s^0} \rightarrow \pi^0 K^0$ &$1.13\times 10^{-8}$ &$1.18\times10^{-7}$
       &$ 7.06\times 10^{-8}$ &$3.36\times10^{-7}$& $2.94\times 10^{-7}$
       &$-6.05\%$ & $-10.2\%$ &$18.5\%$& $23.2\%$ \\
\hline
$\bar{B_s^0} \rightarrow \eta K^0$ &$1.12\times 10^{-8}$ &$1.48\times10^{-6}$
       &$1.62\times 10^{-6}$ &$1.53\times10^{-6}$&$1.67\times 10^{-6}$
       &$5.27\%$ & $4.86\%$ &$6.44\%$  & $5.96\%$ \\
\hline
$\bar{B_s^0} \rightarrow \eta^{'} K^0$ &$3.13\times 10^{-9}$
&$8.33\times10^{-6}$
       &$8.23\times 10^{-6}$ &$8.14\times10^{-6}$&$ 8.04\times 10^{-6}$
       &$3.09\%$ & $3.11\%$ &$2.41\%$  & $2.43\%$ \\
\hline
\end{tabular}
\end{center}

\newpage

\begin{flushleft}
\section*{Figure Captions}
\end{flushleft}

\vspace{1cm}

{\bf Fig. 1.} Quark diagrams for a $B$  meson decaying into two light
pseudoscalar mesons $P$ and $P{'}$ through the tree process
$b\rightarrow u({\bar u}q)$: a) the internal $W$-emission diagram,
(b) the external $W$-emission diagram; and the time-like penguin diagram
process $b\rightarrow q(q{'}{\bar q}{'})$: c) the time-like pure penguin
diagram, and d) the time-like  hairpin diagram. The subscripts ``s" denote
``spectator". The dark dot stands for the contraction of the W-loop.\\
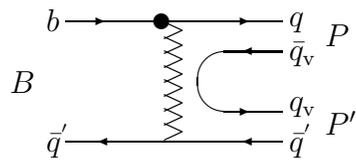
{\bf Fig. 2.} Quark diagrams for a $B$  meson decaying into two light
pseudoscalar mesons $P$ and $P{'}$ through the space-like penguin process
$(b\bar q{'})\rightarrow (q{\bar q}{'})$. The subscripts ``v" denote ``vacuum".
The dark dot stands for the contraction of the W-loop.
\newpage

\unitlength=0.32mm
\begin{figure}

\begin{picture}(350,275)
\put(70,240){\line(1,0){90}}
\put(70,190){\line(1,0){90}}
\put(62,236){$b$}
\put(62,186){$\bar{q}_{\rm s}$}
\put(47,210.5){$B$}
\put(163,238){$u$}
\put(163,186){$\bar{q}_{\rm s}$}
\put(163,224){$\bar u$}
\put(163,203){$q$}
\put(178,193){$P{'}$}
\put(178,230){$P$}
\put(160,215){\oval(70,25)[l]}
\put(85,240){\vector(1,0){2}}
\put(85,190){\vector(-1,0){2}}
\put(145,240){\vector(1,0){2}}
\put(145,190){\vector(-1,0){2}}
\put(145,227.5){\vector(-1,0){2}}
\put(145,202.5){\vector(1,0){2}}
\multiput(110,240)(3,-5){5}{\line(0,-1){5}}
\multiput(107,240)(3,-5){6}{\line(1,0){3}}
\put(110,165){a)}
\put(270,210){\line(1,0){90}}
\put(270,190){\line(1,0){90}}
\put(262,206){$b$}
\put(262,186){$\bar{q}_{\rm s}$}
\put(247,194.5){$B$}
\put(363,208){$u$}
\put(363,186){$\bar{q}_{\rm s}$}
\put(363,248){$q$}
\put(363,228){$\bar u$}
\put(378,194.5){$P{'}$}
\put(378,234.5){$P$}
\put(360,240){\oval(70,20)[l]}
\put(285,210){\vector(1,0){2}}
\put(285,190){\vector(-1,0){2}}
\put(345,210){\vector(1,0){2}}
\put(345,190){\vector(-1,0){2}}
\put(345,230){\vector(-1,0){2}}
\put(345,250){\vector(1,0){2}}
\multiput(310,210)(3,5){5}{\line(0,1){5}}
\multiput(307,210)(3,5){6}{\line(1,0){3}}
\put(310,165){b)}
\end{picture}

\begin{picture}(350,275)
\put(110,240){\circle*{7}}
\put(70,240){\line(1,0){90}}
\put(70,190){\line(1,0){90}}
\put(62,236){$b$}
\put(62,186){$\bar{q}_{\rm s}$}
\put(47,210.5){$B$}
\put(163,238){$q$}
\put(163,186){$\bar{q}_{\rm s}$}
\put(163,224){$\bar{q}^{'}$}
\put(163,203){$q^{'}$}
\put(178,193){$P{'}$}
\put(178,230){$P$}
\put(160,215){\oval(70,25)[l]}
\put(85,240){\vector(1,0){2}}
\put(85,190){\vector(-1,0){2}}
\put(145,240){\vector(1,0){2}}
\put(145,190){\vector(-1,0){2}}
\put(145,227.5){\vector(-1,0){2}}
\put(145,202.5){\vector(1,0){2}}
\multiput(110,240)(3,-5){5}{\line(0,-1){5}}
\multiput(107,240)(3,-5){6}{\line(1,0){3}}
\put(110,165){c)}
\put(310,210){\circle*{7}}
\put(270,210){\line(1,0){90}}
\put(270,190){\line(1,0){90}}
\put(262,206){$b$}
\put(262,186){$\bar{q}_{\rm s}$}
\put(247,194.5){$B$}
\put(363,208){$q$}
\put(363,186){$\bar{q}_{\rm s}$}
\put(363,248){$q^{'}$}
\put(363,228){$\bar{q}^{'}$}
\put(378,194.5){$P{'}$}
\put(378,234.5){$P$}
\put(360,240){\oval(70,20)[l]}
\put(285,210){\vector(1,0){2}}
\put(285,190){\vector(-1,0){2}}
\put(345,210){\vector(1,0){2}}
\put(345,190){\vector(-1,0){2}}
\put(345,230){\vector(-1,0){2}}
\put(345,250){\vector(1,0){2}}
\multiput(310,210)(3,5){5}{\line(0,1){5}}
\multiput(307,210)(3,5){6}{\line(1,0){3}}
\put(310,165){d)}
\put(200,140){Fig. 1.}
\end{picture}

\begin{picture}(350,275)
\put(210,240){\circle*{7}}
\put(170,240){\line(1,0){90}}
\put(170,190){\line(1,0){90}}
\put(162,236){$b$}
\put(162,186){$\bar{q}^{'}$}
\put(147,210.5){$B$}
\put(263,238){$q$}
\put(263,186){$\bar{q}^{'}$}
\put(263,224){$\bar{q}_{\rm v}$}
\put(263,203){$q_{\rm v}$}
\put(278,193){$P{'}$}
\put(278,230){$P$}
\put(260,215){\oval(70,25)[l]}
\put(185,240){\vector(1,0){2}}
\put(185,190){\vector(-1,0){2}}
\put(245,240){\vector(1,0){2}}
\put(245,190){\vector(-1,0){2}}
\put(245,227.5){\vector(-1,0){2}}
\put(245,202.5){\vector(1,0){2}}
\multiput(210,233)(0,-6){8}{$>$}
\put(200, 165){Fig. 2.}
\end{picture}
\caption{}
\end{figure}

\end{document}